# Modeling COVID-19 in Cape Verde Islands - An application of SIR model

Adilson da Silva[*]

*Faculty of Science and Technology*
*University of Cape Verde, Cape Verde*

*Center for Mathematics and Applications*
*New University of Lisbon, Portugal*

**Abstract**

The rapid and surprised emergence of COVID-19, having infected three million and killed two hundred thousand people worldwide in less than five months, has led many experts to focus on simulating its propagation dynamics in order to have an estimated outlook for the not too distante future and so supporting the local and national governments in making decisions. In this paper, we apply the SIR model to simulating the propagation dynamics of COVID-19 on the Cape Verde Islands. It will be done firstly for Santiago and Boavista Islands, ant then for Cape Verde in general. The choice of Santiago rests on the fact that it is the largest island, with more than 50% of the Population of the country, whereas Boavista was chosen because it is the island where the first case of COVID-19 in Cape Verde was diagnosed. Observations made after the date of the simulations were carried out corroborates our projections.

## 1. Introduction

Over several centuries infectious diseases have brought great challenges to humanity, often causing epidemics and pandemics, and claiming thousands up to millions of life. In the same proportion, the worldwide medical class has faced enormous challenges in the search for cures, which usually takes several months and sometimes years to be found. Nowadays, even with all the technological and biotechnological advancement in the medical and pharmaceutical industry of $21^{st}$ century, cures have not yet been found for many diseases; a remarkable and classical example is AIDS (caused by HIV) which has no cure yet even after more than thirty years since the first cases were discovered. Based on estimates, in 2020 the Global Health Observatory (GHO) reported 300 thousand deaths by AIDS per year (see GHO [21]).

[*]Corresponding author
*Email address:* `adilson.dasilva@docente.unicv.edu.cv` (Adilson da Silva )

Chronologically, some of the most devastating infectious diseases in the last seven centuries are summarized below:

- *Black death* (due to the black plague) in Europe, Asia and North Africa claimed up to 72 - 200 million live ($14^{th}$ century, 1331–1353; see Austin and suzanne [23]);

- *New England's (coastal) Plague* claimed up to 30-90% of the population ($17^{th}$ century, 1616–1619; see Bratton [24]);

- *Persian plague* claimed 2 million lives ($18^{th}$ century, 1772–1773; see Shahraki et al. [25]);

- *Third cholera pandemic* in Russia claimed over one million lives ($19^{th}$ century, 1852-1860; see Hays [26]);

- *Influenza pandemic of 1918* (the most severe pandemic in recent history) claimed at least 50 million lives worldwide ($20^{th}$ century, 1918–1919; see Patterson and Pyle [27]);

- *Asian flu* (also known as 1957-58 Influenza pandemic) claimed two million lives worldwide ($20^{th}$ century, 1957–1969; see Paul [28]);

- *HIV* (responsible for AIDS) claimed over 30 million lives worldwide and still claiming lives, since there is no cure yet (XX century, since 1960; see Ortiz et al. [29]).

Early in the $21^{st}$ century, specifically in 2009, the world has witnessed the pandemic of flu A of 2009 caused by H1N1 influenza virus , claiming more than 18 thousand lives worldwide (see World Health Organization [22]). Between 2013 and 2016 we also witnessed the ebola outbreak in West Africa which pinpointed serious flaws in emergency response systems, since it ultimately claimed 11 thousand of lives by the time it was officially declared over (see Baseler et al. [30]).

As we write, we are faced white COVID-19 pandemic (acute respiratory disease caused by novel coronavirus), which started by the end of 2019 in China, and reached almost every country in the world with devastating results. Unfortunately, within five months since the first case was reported, by April 24 the World Health Organization reported more than 2 million 800 thousand infected people, 195 thousand deaths, and 776 thousand recovered cases. While researchers struggle to find the cure for this devastating infectious disease, the numbers of infected and death cases continue to increase exponentially; however, fortunately, at the same time mathematicians from several countries rush to model its propagation dynamics by making projections for months ahead, in order to support national government authorities in the control measures.

This study aims to make projection of the propagation dynamics of COVID-19 in Cape Verde (see Figure 1). It is organized as follow: the next two subsections, 1.1 and 1.2, introduce the materials and model used; Section 2 introduces and discusses the model and the methodology adopted; Section 3 is devoted to the projection of COVID-19 in Cape Verde for the next 3 to 6 months; finally, Section 4 discusses the achieved results.

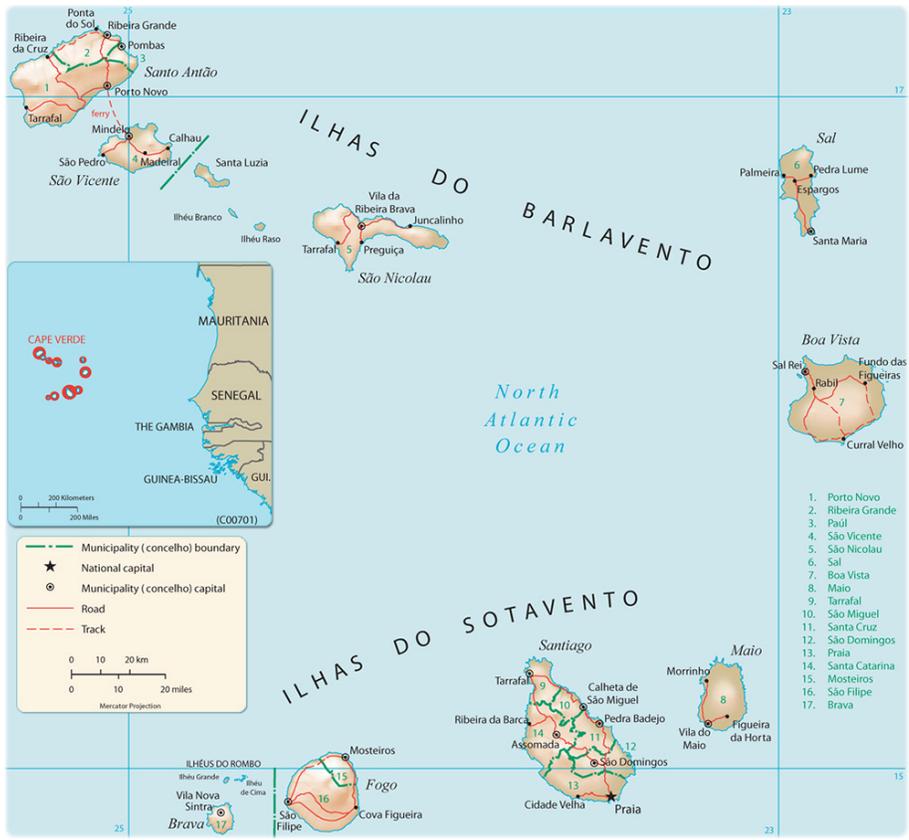

Figure 1: Map of Cape Verde (see the African Geographic Guide [31])

### 1.1. Materials

Here we introduce the proposed mathematical model for COVID-19 analysis on the Cape Verde Islands. The model parameters are estimated via optimization of the error function given as square difference between the number of infected and the corresponding number of predicted cases by the model we are using. The data are selected accordingly with the occurrence of the first infected case in Santiago (2020-03-25), Boavista (2020-03-19) and Cape Verde in general (2020-03-19). Starting on 2020-03-19 or 2020-03-25, as the case may be, the observations were made up until 2020-04-17. The date follows the ISO 8601format YYYY-MM-DD.

*1.2. Proposed Model*

The epidemiological models divide the target population into compartments reflecting the different stages in which the individuals find themselves during a disease propagation at time $t \in [0, +\infty[$, such as "Immune", " Susceptible", "Infected', "Quarantine", "Removed", "Recovered", "Death", etc, whose evolution in time is described by continuous functions, and model the relations between the compartments with ordinary differential equations involving the relative functions.

During the $20^{th}$ century several researchers have conducted researches in epidemic modeling (see Herbert [1], [3], Metz and Van den Bosch [5] and Diekmann et al. [13], Li et al. [14], Chowell et al. [20], Schuette [8], Mollisson [6], for example), in an effort to allow a better understanding of the propagation dynamics and the disease impact in order to support government authorities in the control measures.

Kermack and Mackendrick [18] (see also Isea and Lonngren [19]) proposed the one that would become the epidemic model playing central role, the SIR model; it is known as compartmental model due to the fact that it divides the population into three compartments:

- Susceptive (S): individuals susceptible to be infected;

- Infected (I): Infected individuals ;

- Removed (R): Individuals removed from compartment I (recovered or dead).

SIR model has been used to model epidemic diseases (see Busenberg and Cooke [12], for example) such as the black plague in Europe, Asia and North Africa, New England epidemic, Persian plague and Spanish flu of 1915, and has produced accurate projections (see Herbert [1] and [3]). It must be remarked that it makes no distinction between those who recovered and those who died, all of them are put into compartment R. It is also relevant to remark that the recovered individuals are assumed to acquire immunities at least within the period in which the studies is conducted. This is a reasonable assumptions, mostly because those who have recovered from an infectious disease will not expose themselves again so easily. SIR model also does not take into account the latent period, it is assumed that as soon as an individual is infected he becomes a vehicle to infect other individuals.

Let S(t), I(t) and R(t) respectively denote the number of individuals at compartments of Susceptible, Infectious and Removed, at time $t \in [0, +\infty[$, and assumes that the population is homogeneous and there is no vital dynamics, that is the total number of individuals remain equal to N, where $N = S(t)+I(t)+R(t)$ at any time $t$. It is also assumed that time $t$ is counted in days, that is $t \in [0, 1, 2, \dots[$.

## 2. Background - SIR Model

Under the hypothesis previously considered, the dynamics of disease propagation via contact may be modeled with SIR model accordingly. Let $\beta$ denotes the parameter that control the transition between S and I, called *contact rate*, and $\gamma$ the one which control the transition between I and $R$, called *remotion rate* (rate with which an individual died or recovered from the disease). Figure 2 shows the flow diagram of the approached model.

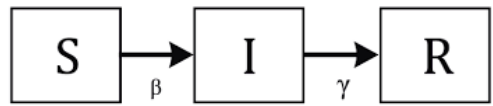

Figure 2: A flow diagram demonstrating the flux between the three compartments $S$, $I$ and $R$, as well as the rate $\beta$ and $\gamma$ at which the individuals move between the compartments.

The SIR model is given as the following system of nonlinear ordinary differential equations:

$$\begin{cases} S'(t) = -\left(\frac{\beta}{N}\right)S(t)I(t) \\ I'(t) = \left(\frac{\beta}{N}\right)S(t)I(t) - \gamma I \\ R'(t) = \gamma I(t) \\ N = S(t) + I(t) + R(t), \ t \in [0, +\infty[ \\ S(0) = S_0 > 0, I(0) = I_0 > 0, R(0) = R_0 \geq 0 \text{ (conditions at time t=0)}, \end{cases} \quad (1)$$

where $S'(t)$ denotes the derivate of function $S$ at $t$, that is $\frac{dS(t)}{dt}$.

This model uses the standard incidence and has recovery rate at time $t$ as $\gamma I(t)$, corresponding to an exponential waiting time $e^{-\gamma t}$, and since the time period (day) is short it does not take into account the vital dynamics (births and deaths).

Let $s(t) = \frac{S(t)}{N}$, $i(t) = \frac{I(t)}{N}$ and $r(t) = \frac{R(t)}{N}$; this yields $s(t) + i(t) + r(t) = 1$.

The system (1) can be rewritten as follows:

$$\begin{cases} s'(t) = -\beta s(t)i(t) \\ i'(t) = \beta s(t)i(t) - \gamma i(t) \\ r'(t) = \gamma i(t) \\ 1 = s(t) + i(t) + r(t), \ t \in [0, +\infty[ \\ s(0) = s_0, i(0) = i_0, r(0) = r_0 \text{ (conditions at time t=0)}, \end{cases} \quad (2)$$

Since $r(t)$ can always be found from $s(t)$ and $i(t)$ by using the fourth equation of system (2), it is sufficient to consider (2) in the plane $si$, i.e.:

$$\begin{cases} s'(t) = -\beta s(t)i(t) \\ i'(t) = \beta s(t)i(t) - \gamma i(t) \\ s(t) + i(t) \leq 1, \ t \in [0, +\infty[ \\ s_0 > 0, i_0 > 0 \end{cases} \quad (3)$$

The system (3) shows that the epidemiologically region is the triangle T given by

$$T = \{(s(t), i(t)) : s(t) \geq 0, i(t) \geq 0, s(t) + i(t) \leq 1\}.$$

.

**Theorem 2.1.** *Let (s(t),i(t)) be the solution of (3) in T.*

(a) *If $\frac{\beta s_0}{\gamma} \leq 1$, then $i(t) \to 0$ as $t \to +\infty$;*

(b) *If $\frac{\beta s_0}{\gamma} > 1$, then $i(t)$ first increases up to a maximum value*

$$i_{max} = s_0 + i_0 - \frac{\gamma}{\beta} - \frac{\gamma}{\beta} log\left(\frac{\beta s_0}{\gamma}\right)$$

*and then decreases to 0 as $t \to 0$, where log denotes the natural logarithm;*

(c) *s(t) is a decreasing function of t and has an horizontal assimptota $s_* = lim_{t \to +\infty} s(t)$, where $s_*$ is the unique root of the equation*

$$s_0 + i_0 - s_* + \frac{\gamma}{\beta} log\left(\frac{s_*}{s_0}\right) = 0$$

*in $\left]0, \frac{\gamma}{\beta}\right[$.*

*Proof.* See the Appendix of Herbert [1]. □

The threshold $\frac{\beta s_0}{\gamma}$ in the Theorem 2.1, known as $R_o$, is a quantity that measure how fast the propagation of a disease is. We may note that at the beginning of an epidemic, day 1 (t=0), we have that $i_0$ is a very small value, usually equal one, and therefore

$$s_0 = \frac{S_0}{N} = \frac{N - i_0}{N} \approx 1,$$

since the population is usually large. Thus, $\beta s_0$ is the number of susceptible individuals that are infected by contact with one infected individual per day at beginning of the epidemic. On the other hand, the quantity $\gamma^{-1}$ is the average duration of infection, that is the average period in which an individual remains infectious.

**Definition 2.1.** $\mathcal{R}_o = \frac{\beta s_0}{\gamma}$ *is the basic reproduction number per unity of time (day).*

In other words $\mathcal{R}_o$ is the number of individuals infected by one infected individual per day. See Diekmann et al. [13] or Chowell et al. [20] for additional information.

## 3. Simulations

Let G(t)=(S(t), I(t), R(t)), depending on the parameters $\beta$ and $\gamma$, where $0 \leq \beta, \gamma \leq 1$, denoting the unknown solution of SIR model for the number of Susceptible, Infected and Removed individuals, and $A(t) = (S^a(t), I^a(t), R^a(t))$ denoting the corresponding actual number of observed Susceptive, Infected and Removed individuals in Cape Verde at time (day) $t$. Since we also aim to project the propagation dynamics of disease on Santiago and Boavista islands, it is convenient for us to introduce the corresponding notations, that is $S_S(t)$, $I_S(t)$ and $R_S(t)$ respectively denote the number of Susceptive, Infected and Removed individuals for Santiago island, while $S_B(t)$, $I_B(t)$ and $R_B(t)$ the corresponding numbers for Boavista island. More over, according to the corresponding context, Cape Verde, Santiago, and Boavista, the following notation will be used without additional explanation:

- $G_S(t) = (S_S(t), I_S(t), R_S(t))$ and $G_B(t) = (S_B(t), I_B(t), R_B(t))$ respectively denote the unknown solution of SIR model for the number of Susceptible, Infected and Removed individuals in Santiago and Boavista.

- $A_S^a(t) = (S_S^a(t), I_S^a(t), R_S^a(t))$ and $A_B^a(t) = (S^B(t), I^B(t), R^B(t))$ respectively denote the number of observed Susceptive, Infected and Removed individuals in Santiago and Boavista;

- $N$, $N_S$ and $N_B$ respectively denote the constant number of population of Cape Verde, Santiago and Boavista;

- $\beta$, $\beta_S$ and $\beta_b$ respectively denote the *contact rate* for Cape Verde, Santiago and Boavista;

- $\gamma$, $\gamma_S$ and $\gamma_B$ respectively denote the *contact rate* for Cape Verde, Santiago and Boavista;

- $\|.\|_2$ denotes the Euclidean norm.

In order to find the optimal estimates for the parameters $(\beta, \gamma)$, $(\beta_S, \gamma_S)$ and $(\beta_B, \gamma_B)$ for which the SIR model better fit the observed date from 2020-03-25

to 2020-04-17 in Santiago, from 2020-03-19 to 2020-04-17 in Boavista, and from 2020-03-19 to 2020-04-17 in Cape Verde, respectively, we use the L-BFGS-B algorithm (see ScipOrg [32]) through the *optim*() function from R software to solve the least-square problems:

$$\underset{0\leq\beta,\gamma\leq 1,\ 1\leq t\leq 29}{\operatorname{argmin}} \|G(t) - A^a(t)\|_2^2; \quad (4)$$

$$\underset{0\leq\beta_S,\gamma_S\leq 1,\ 1\leq t\leq 24}{\operatorname{argmin}} \|G_S(t) - A_S^a(t)\|_2^2; \quad (5)$$

$$\underset{0\leq\beta_B,\gamma_B\leq 1,\ 1\leq t\leq 29}{\operatorname{argmin}} \|G_B(t) - A_B^a(t)\|_2^2. \quad (6)$$

*3.1. Projection for Santiago and Boavista Islands*

In this Section we propose and discuss the projection of COVID-19 for Santiago and Boavista. It will be done firstly for Santiago (subsection 3.1.1) and then for Boavista (subsection 3.1.2).

*3.1.1. Santiago Insland*

With a total of $N_S = 313461$ inhabitants, being 170236 of them located in the city of Praia, the largest city of Cape Verde, Santiago is home to more the $\frac{1}{2}$ of Cape Verde's total population; this makes Santiago - largest island in territory and population - the island which requires more attention with regards to the propagation of an infectious disease like COVID-19.

The records for 24 days (from 2020-03-25 to 2020-04-25) of observations in Santiago are summarized in Table 1.

Table 1: Records from 25/03/2020 to 17/04/2020 in Santiago. For of values $S_S(t)$, $t = 1, \ldots, 24$, it is enough to recall that $S_S(t) = N_S - I_S(t) - R_S(t)$.

| | Data from 24 days of observations in Santiago: 25/03/2020 to 17/04/2020. | | | | | | | | | | | | | | | | | | | | | | | |
|---|---|---|---|---|---|---|---|---|---|---|---|---|---|---|---|---|---|---|---|---|---|---|---|---|
| Days (t) | 1 | 2 | 3 | 4 | 5 | 6 | 7 | 8 | 9 | 10 | 11 | 12 | 13 | 14 | 15 | 16 | 17 | 18 | 19 | 20 | 21 | 22 | 23 | 24 |
| Infected | 1 | 1 | 0 | 0 | 0 | 0 | 0 | 0 | 0 | 0 | 0 | 0 | 0 | 0 | 0 | 0 | 1 | 0 | 0 | 0 | 0 | 0 | 0 | 1 |
| $I_S^a(t)$ | 1 | 2 | 2 | 2 | 2 | 2 | 2 | 2 | 2 | 2 | 2 | 2 | 2 | 2 | 2 | 2 | 3 | 3 | 3 | 3 | 3 | 3 | 3 | 4 |
| $R_S^a(t)$ | 0 | 0 | 0 | 0 | 0 | 0 | 0 | 0 | 0 | 0 | 0 | 1 | 1 | 1 | 1 | 1 | 1 | 1 | 1 | 1 | 1 | 1 | 1 | 1 |

Through application of the *optim*() function to solve the least-square problem (5), regarding the observed data (see Table 1), the optimal estimates for parameters $\beta_S$ and $\gamma_S$ was produced and the results summarized in Table 2.

Table 2: Estimates for parameters $\beta_S$, $\gamma_S$, $s_0^S$ and $\mathcal{R}_o^S$ from the observed data.

| Contact rate | Remot. rate | Prop. Suscept. | Repr. number |
| --- | --- | --- | --- |
| $\beta$ | $\gamma$ | $s_0^S$ | $\mathcal{R}_o^S$ |
| 0.7555025 | 0.6977990 | 0.9999968 | 1.08269 |

The $\mathcal{R}_o^S = 1.08269$ shows that the velocity of the propagation in Santiago is largely below the range $]1.5, \ 3.5[$, which is the observed range for the European, Asiatic and American countries. This means that each infected individual transmits the infection to 1.08 individual per day. In addition, this small value for $\mathcal{R}_o^S$ somehow reflects the correct measures adopted by the government early after the first case of infection emerged in Cape Verde, precisely at Boavista island. It is known that as soon as the concerned disease spread across Europe, the Cape Verde government closed the air connections with all the affected countries, Italy being the first one, and recommended social distancing within the country. Three days after the first case in Santiago was recorded the government declared mandatory isolation and closed all internal air and sea connections, only allowing individuals to leave home for strictly regulated basic necessities. The inter-island connections was (and still are as we write) also allowed only for transporting supplies but even so, under strict regulation. All these measures seem to have influenced the value of $\mathcal{R}_o^S$.

Now we use the achieved optimal estimates for *contact* and *remotion* rates to predict the propagation of COVID-19 in Santiago. It will be done by using the *ode*() function from the R package "deSolve" (see Soetaert et al. [33]) to solve the differential equations of SIR model. The produced results are summarized in Figures 3 and 4; Figure 4 is the same as Figures 3 but with the estimated peak of infected individuals highlighted; the estimated number for total deaths is also highlighted in Figure 4.

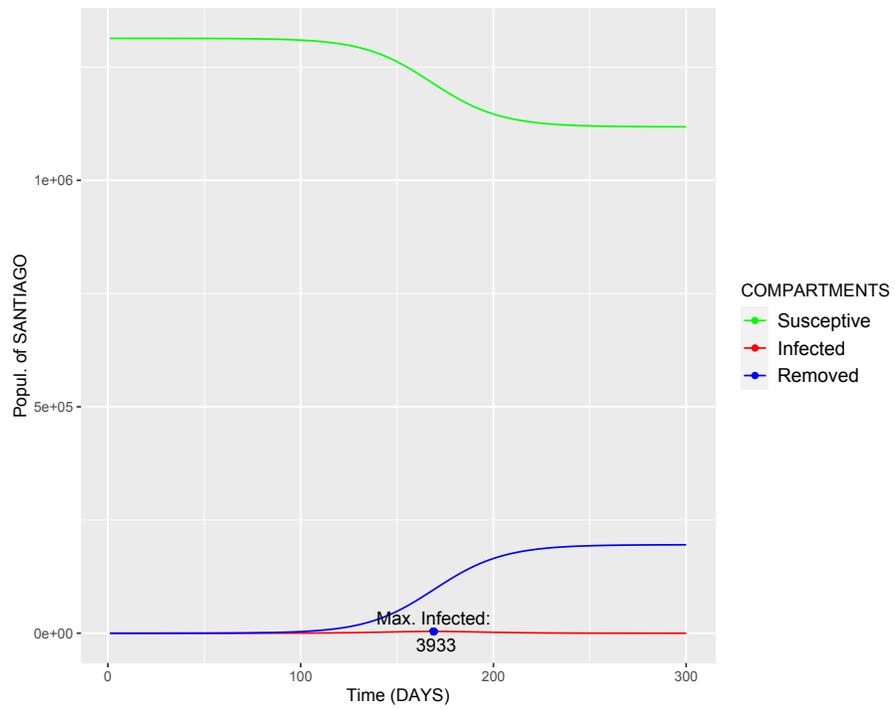

Figure 3: Projection of COVID-19 in Santiago, using data from 2020-03-25 to 2020-04-17.

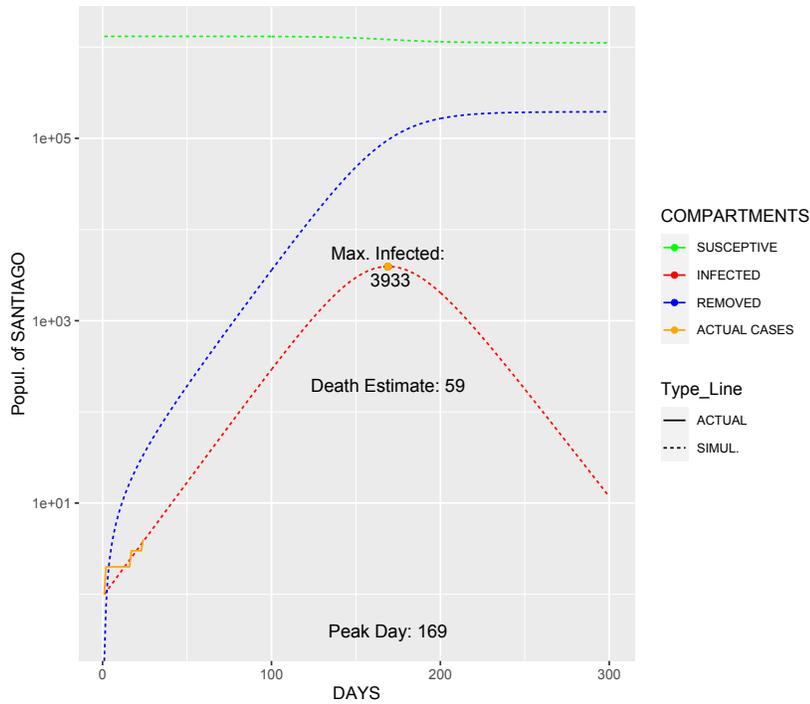

Figure 4: Projection of COVID-19 in Santiago, using data from 2020-03-25 to 2020-04-17.

Figures 3 and 4 shows that the propagation will be increasing for 169 days, reaching a peak of 3933 infected, and at that point it starts to decrease until disappear from the Santiago. This results are achieved on the base that the island remains isolated.

*3.1.2. Boavista Island*

On its turn, Boavista is small both in territory and population compared to Santiago; its total population of $N_B = 19879$ is $\frac{1}{15}$ of Santiago's population. The first and main reason for studying the propagation of COVID-19 on this island, and not on the rest of the other seven Islands (except Santiago), is the fact that the first reported case in Cape Verde (on 2020-03-19) took place there. Another reason is the fact that the reported case was related to a foreign tourist hosted in Hotel "Riu Karamboa", which is one of the biggest hotels in Cape Verde with more than 500 rooms. The last reason is the fact that Boavista's main activities are all related to tourism, having more than 25 hotels for a small population of $N_B = 19879$ habitants, and almost all local family income sources are related to some contractual positions with local hotels.

All the above-mentioned situations are extremely conducive to a massive dissemination of the disease in a short period of time, and so our particular interest in studying the propagation there.

The records for 29 days (from 2020-03-19 to 2020-04-17) of observations in Boavista are summarized in Table 3.

Table 3: Records from 2020-03-19 to 2020-04-17 in Boavista. For of values $S_S(t)$, $t = 1, \ldots, 29$, it is enough to recall that $S_S(t) = N_S - I_S(t) - R_S(t)$

| Data from 29 days of observations in Boavista: 2020-03-19 to 2020-04-17 |
|---|
| Days (t) | 1 | 2 | 3 | 4 | 5 | 6 | 7 | 8 | 9 | 10 | 11 | 12 | 13 | 14 | 15 | 16 | 17 | 18 | 19 | 20 | 21 | 22 | 23 | 24 | 25 | 26 | 27 | 28 | 29 |
| Infected | 1 | 2 | 0 | 0 | 0 | 0 | 0 | 0 | 1 | 0 | 0 | 0 | 0 | 0 | 0 | 0 | 0 | 0 | 0 | 0 | 0 | 0 | 0 | 2 | 0 | 1 | 44 | 0 | 0 |
| $I_B^o(t)$ | 1 | 3 | 3 | 3 | 3 | 3 | 3 | 3 | 4 | 4 | 4 | 4 | 4 | 4 | 4 | 4 | 4 | 4 | 4 | 4 | 4 | 4 | 4 | 6 | 6 | 7 | 51 | 51 | 51 |
| $R_B^o(t)$ | 0 | 0 | 0 | 0 | 1 | 1 | 1 | 1 | 1 | 1 | 1 | 1 | 1 | 1 | 1 | 1 | 1 | 1 | 1 | 1 | 1 | 1 | 1 | 1 | 1 | 1 | 1 | 1 | 1 |

As stated for the previous case, the *optim*() function is used to solve the least-square problem (6), regarding the observed data (see Table 3), the optimal estimates for parameters $\beta_B$ and $\gamma_B$ was produced and the results summarized in Table 4.

Table 4: Estimates for parameters $\beta_B$, $\gamma_B$, $s_0^B$ and $\mathcal{R}_o^B$ from the observed data

| Contact rate | Remotion rate | Prop. Suscept. | Reprod. number |
|---|---|---|---|
| $\beta_B$ | $\gamma_B$ | $s_0^B$ | $\mathcal{R}_o^B$ |
| 0.4811817 | 0.3500000 | 0.9999497 | 1.374736 |

As it may be seen, $\mathcal{R}_o^B = 1.374736$, which is relatively largely than $\mathcal{R}_o^S$. This, therefore, indicates that the disease is spreading relatively fast in Boavista than in Santiago, but yet still below the observed range in Europe, Asia and America. This value seems to be largely related to the on time above-mentioned government measures. This value for $\mathcal{R}_o^B$ suggests that, in average, each infected individual transmits the infection to 1.37 individuals per day. Next, the achieved optimal estimates for *contact* and *remotion* rates will be used to predict the propagation of COVID-19 in Boavista. It will be done by using the $ode()$ function to solve the differential equations of SIR model. The produced results are summarized in Figures 5 and 6; Figure 6 is the same as Figures 5 but with the estimated peak of infected individuals highlighted. The estimated number for total deaths is also highlighted in Figure 6.

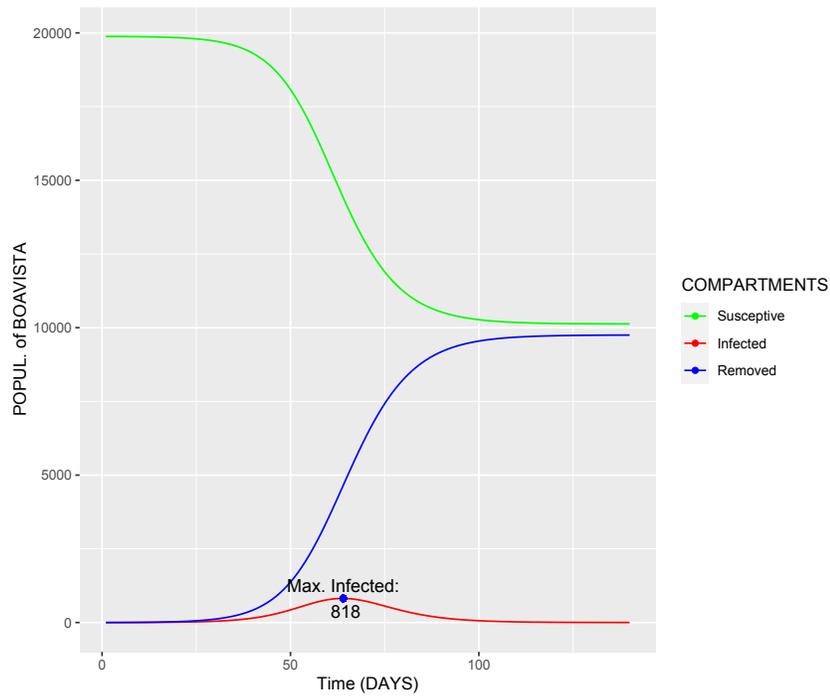

Figure 5: Projection of COVID-19 in Boavista, using data from 2020-03-19 to 2020-04-17.

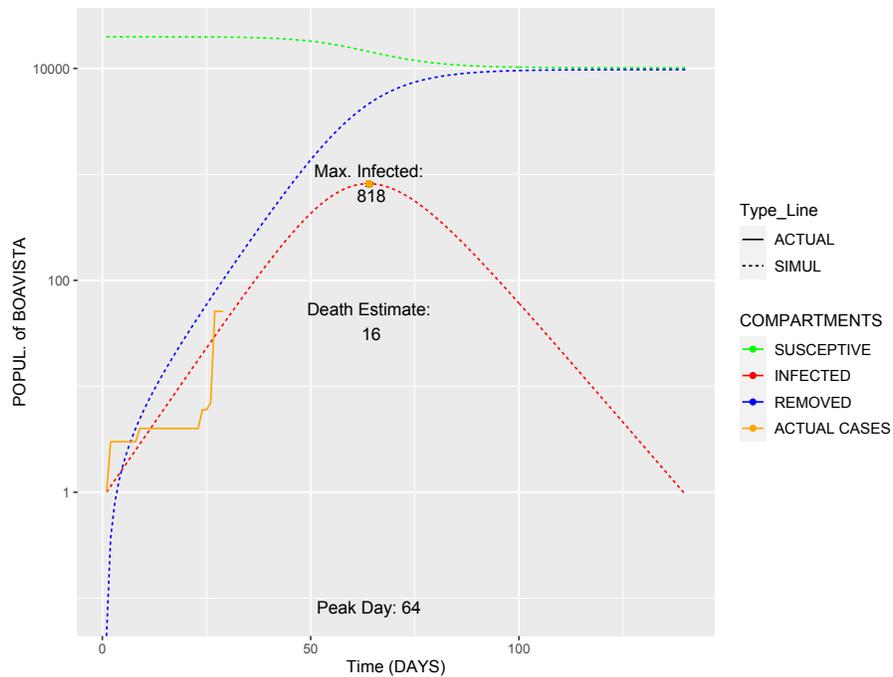

Figure 6: Projection of COVID-19 in Boavista, using data from 2020-03-19 to 2020-04-17.

Figures 3 and 4 suggest that the propagation will be increasing for 64 days, reaching a peak of 818 infected people, and at that point it starts to decrease until it disappears in Boa Vista. This results also was achieved on the basis that the island remains isolated.

### 3.1.3. Cabo Verde

Now we turn to the Cape Verde in general. As previously pointed out, as soon as the first case was reported the government strongly recommended social distancing, while cutting the air and sea connections with the affected island, firstly with Boavista followed by Santiago and S. Vicente, before the total lock down on 2020-03-29. Unfortunately, even with the mandatory lock down the government has been reporting cases in which some individuals were using small boats to travel between the islands, mostly leaving from Boavista which was the island with the highest risk.

The records for 29 days (from 19/03/2020 to 17/04/2020) of observations in Cape Verde are summarized in Table 5. Recall that the total population of Cape Verde is $N = 556586$.

Table 5: Records from 19/03/2020 to 17/04/2020 in Cape Verde. For of values $S(t)$, $t = 1, \ldots, 29$, it is enough to recall that $S(t) = N - I(t) - R(t)$.

| Data from 29 days of observations in Cape Verde: 19/03/2020 to 17/04/2020. | | | | | | | | | | | | | | | | | | | | | | | | | | | | | |
|---|---|---|---|---|---|---|---|---|---|---|---|---|---|---|---|---|---|---|---|---|---|---|---|---|---|---|---|---|---|
| Days (t) | 1 | 2 | 3 | 4 | 5 | 6 | 7 | 8 | 9 | 10 | 11 | 12 | 13 | 14 | 15 | 16 | 17 | 18 | 19 | 20 | 21 | 22 | 23 | 24 | 25 | 26 | 27 | 28 | 29 |
| Infected | 1 | 2 | 0 | 0 | 0 | 0 | 1 | 1 | 0 | 1 | 0 | 0 | 0 | 0 | 0 | 1 | 0 | 0 | 0 | 0 | 0 | 1 | 0 | 2 | 0 | 1 | 44 | 0 | 1 |
| $I^a(t)$ | 1 | 3 | 3 | 3 | 3 | 3 | 4 | 5 | 5 | 6 | 6 | 6 | 6 | 6 | 6 | 7 | 7 | 7 | 7 | 7 | 7 | 8 | 8 | 10 | 10 | 11 | 55 | 55 | 56 |
| $R^a(t)$ | 0 | 0 | 0 | 0 | 1 | 1 | 1 | 1 | 1 | 1 | 1 | 2 | 2 | 2 | 2 | 2 | 2 | 2 | 2 | 2 | 2 | 2 | 2 | 2 | 2 | 2 | 2 | 2 | 2 |

Applying the *optim*() function to the least-square problem (4) regarding the observed data at Table 5, the optimal estimates for parameters $\beta$ and $\gamma$ was produced and the results summarized in Table 6.

Table 6: Estimates for parameters $\beta$ and $\gamma$ from the observed data.

| Contact rate | Remotion rate | Prop. Suscep. | Reprod. Number |
|---|---|---|---|
| $\beta$ | $\gamma$ | $s_0$ | $\mathcal{R}_o$ |
| 0.87 | 0.77 | 0.9999982 | 1.129868 |

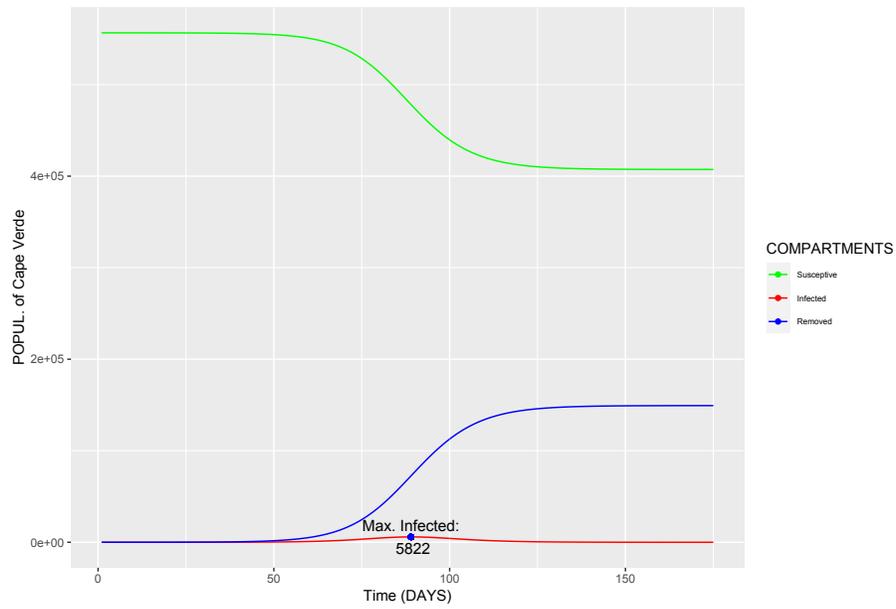

Figure 7: Projection of COVID-19 in Cape Verde, using data from 2020-03-19 to 2020-04-17.

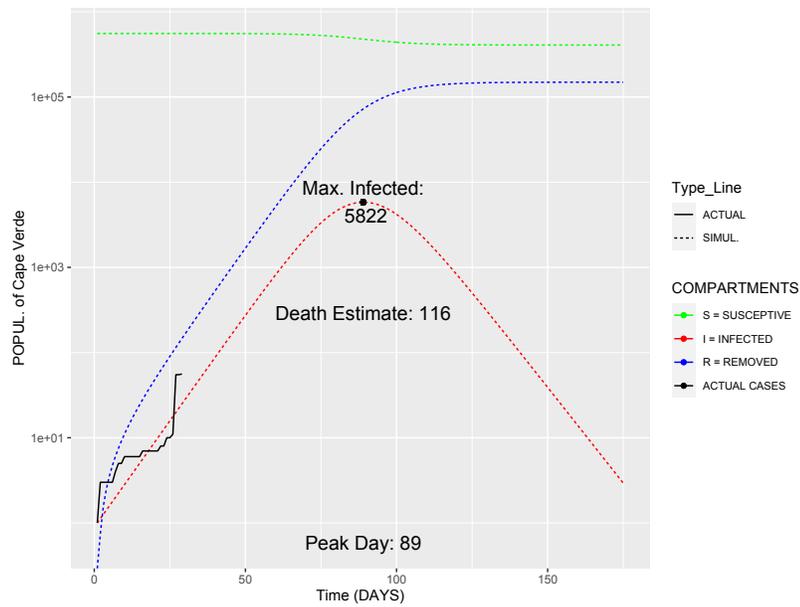

Figure 8: Projection of COVID-19 in Cape Verde, using data from 2020-03-19 to 2020-04-17.

Figures 3 and 4 suggest that the propagation will be increasing for 89 days, reaching a peak of 5822 infected, and at that point it starts to decrease until it disappears from the Cape Verde. The estimated number for total deaths is highlighted in Figure 8.

## 4. Discussion

The Figures 3, 4, 5, 6, 7 and 8 corroborate what is the expected trend of an highly infectious disease like COVID-19 which has already infected almost 3 million people worldwide in less than five months. They show that the number of infectious individuals increases up exponentially (notice that there is exponentially fast and exponentially slow) until a peak (maximum value), and then decreases to zero, this was already ensured by Theorem 2.1 ($b$).

While producing this work, between 2020-04-17 and 2020-04-23, an increase was reported in infectious number somehow consistent with our achievement, although slightly below; this may have been due to issues related to one of the following reasons:

- An extraordinary performance of individuals with the accomplishment of the social distance measures;

- Some defect in conducting tests for disease screening, such as a limitation on the number of tests and the choice of class of individuals for whom the test are conducted. Certainly, the greater the number of individuals screened for the disease screening the less is the probability of having individuals with the disease without being detected.

Since the day the first case was reported up to 2020-04-22, the authorities reported having conducted tests for COVID-19 screening only 342 individuals in the country, yet only in individuals with symptoms or those who were confirmed to be in direct contact with infected individuals, which reveals to be extremely limited in what concerns both the number of individuals and who the test is applied to. In order to reduce as much as possible the possibility of having not-detected (unidentified) infected individuals spreading the disease, we strongly recommend increasing the number of tests as much as possible, and extend it to other classes such as old people, individuals with disability or comorbidities and health professionals on the ground. We also recommend randomly screening individuals in society. Besides all this, as a complement, we strongly recommend to continue following all the social distance measures at least up to 35 - 45 days, and mitigate them as the number of occurrence decreases.

**Acknowledgment**

Many thanks to Prof. Dr. Manuel L. Esquível for the fruitful discussion while doing this work and Prof. Dr. Saidu Bangura for the critical review of the text.